\documentclass{aa}

\usepackage{graphics}
\usepackage{psfig}

\begin{document}
\newcommand{\msun}  {$M_{\odot}$}
\newcommand{\mlt}   {$\alpha_{\rm MLT}$}
\newcommand{\mlta}   {$\alpha_{\rm MLT, A}$}
\newcommand{\mltb}   {$\alpha_{\rm MLT, B}$}
\newcommand{\teff} {$T_{\rm eff}$}
\newcommand{\teffa} {$T_{\rm eff, A}$}
\newcommand{\teffb} {$T_{\rm eff, B}$}
\newcommand{\al}    {\rm et al.}
\newcommand{\met}   {metallicity }
\newcommand{\kms}   {km s$^{-1}$}
\newcommand\beqa{\begin{eqnarray}}
\newcommand\eeqa{\end{eqnarray}}

\title{Disentangling discrepancies between stellar evolution \\
theory and sub-solar mass stars}
\subtitle{The influence of the mixing length parameter for the UV Psc binary}

\author{E. Lastennet \inst{1}, 
        J. Fernandes \inst{1}, 
        D. Valls-Gabaud \inst{2},
        E. Oblak \inst{3}
       }
 
\institute{Observat\'orio Astron\'omico da Universidade de Coimbra, 
	   Santa Clara, P-3040 Coimbra, Portugal
\and
	   UMR CNRS 5572, Laboratoire d'Astrophysique, 
           Observatoire Midi-Pyr\'en\'ees, 
           14, Avenue Edouard Belin, 31400 Toulouse, France
\and
           UMR CNRS 6091, Laboratoire d'Astrophysique, 
	   Observatoire de Besan\c{c}on,  
           41 bis, avenue de l'Observatoire, BP 1615, 
           25010 Besan\c{c}on, France
           }
\offprints{J. Fernandes}

\date{Received  / Accepted }

\authorrunning{E. Lastennet {\al}}
\titlerunning{The UV Psc binary system} 

\abstract{Serious discrepancies have recently been observed between 
predictions of stellar evolution models in the 0.7-1.1 
{\msun} mass range and accurately measured properties of 
binary stars with components in this mass range. We study one of 
these objects, the eclipsing binary UV Piscium, which is particularly  
interesting because Popper (1997) derived age estimates for each component 
which differed by more than a factor of two. In an attempt to solve this 
significant discrepancy (a difference in age of 11 Gyr), we compute a 
large grid of stellar evolution models with the CESAM code for each component. 
By fixing the masses to their accurately determined values (relative error
smaller than 1\% for both stars), we consider a wide range  of possible 
metallicities $Z$ (0.01 to  0.05), and Helium content $Y$ (0.25 to 0.34) 
uncorrelated to $Z$. In addition, the mixing length parameter {\mlt} is 
left as another free parameter. We obtain a best fit in 
the {\teff}-radius diagram for a common chemical composition 
($Z$, $Y$)$=$(0.012, 0.31), but a different MLT parameter  
{\mlta}$=$ 0.95$\pm$0.12(statistical)+0.30(systematic) and 
{\mltb}$=$ 0.65$\pm$0.07(stat)+0.10(syst). 
The apparent age discrepancy found by Popper (1997) disappears 
with this solution, the components being coeval to within 1\%.
This suggests that fixing {\mlt} to its solar value ($\sim$1.6), a common  
hypothesis assumed in most  stellar evolutionary models, may not be correct. 
Secondly, since  {\mlt} is smaller for the less massive component, this 
suggests that the {\mlt} parameter may decrease with stellar mass,
showing yet another shortcoming of the mixing length theory to explain
stellar convection. This trend needs further confirmation with other 
binary stars with accurate data. 
}

\maketitle

\section{Introduction}

It is of the utmost importance for stellar evolution theory to match at least
the best known objects (the Sun and non-interacting binary systems)
before any attempt to derive proper information for star clusters or
stellar populations in galaxies is made. 

Some binary stars are known to provide uncomparable astrophysical laboratories 
to calibrate theoretical stellar evolutionary models. 
Since the individual components of well-detached binary systems can be assumed 
to be two single-like stars with a common origin, they share the
same chemical composition and same age, and, therefore, the observed 
parameters 
of both components are expected to be matched by a single isochrone at the 
same chemical composition. 

Whilst in general a good overall agreement is found, some studies
(pioneered by  Popper (1997), see also  
e.g. Pols {\al} (1997), Lastennet {\al} (1999), Clausen {\al} (1999) and 
Lastennet \& Valls-Gabaud (2002) for a systematic analysis) have 
unambiguously found that systems with components in the 0.7 to 1.1 \msun
~mass range raise serious difficulties. 
 
There are many well-studied systems with components in this mass
range which
present puzzling discrepancies with the predictions of stellar evolution
theory. For example, if the seismic observations of $\alpha$ Cen A are fitted
to state-of-the-art structure models, the resulting masses differ from the
ones derived from dynamical analyses (Pourbaix {\al} 2002, Th\'evenin
{\al} 2002). The visual binary 85 Peg, some 
eclipsing binaries in the field, and in the Hyades open cluster 
are other clear examples (see  Lastennet {\al} (2002) for 
a brief review of the results obtained so far on these objects). 

This situation is very puzzling because one would expect these stars
to be reasonably well understood. Indeed,    
stars with masses larger than $\sim$0.6 M$_{\odot}$
bypass difficulties in the treatment of the equation of state and
the atmosphere, while
stars with masses larger than about 1.1 M$_{\odot}$ have a permanent convective
core, introducing an additional parameter, the amount of overshooting,
for their modelling. 
Hence, current stellar evolutionary models are expected to be able to match 
the basic properties of stars in the 0.7 -- 1.1 M$_{\odot}$ mass range, 
provided a good description of the convection in their envelopes is used.  

One of these puzzling objects is the eclipsing binary UV Piscium 
(HD 7700, HIP 5980, hereafter UV Psc), a detached main-sequence binary 
(with components 
of types G and K), which is also a short period ($<$1 day)
 RS CVn\footnote{RS CVn are binary systems where one of the components
presents an extreme form of solar-like activity with starspots and variable
magnetic fields. See Montesinos {\al} (1988) for a discussion. }.
This binary deserves particular attention ever since Popper (1997) 
derived age estimates for each component which differed 
by a factor larger than two, with an absolute age difference larger 
than 11 Gyr.
Montesinos {\al} (1988) classified its components as ``apparently normal
main-sequence stars''. What is the origin of this discrepancy?

Popper (1997) used the Geneva models (Schaller {\al} 1992) for $Z=$0.02 and 
$Y=$0.30, i.e. with a helium abundance to metallicity  ratio fixed to 
$\Delta$$Y$/$\Delta$$Z$$=$3, and {\mlt} 
 fixed to a solar-calibrated value ({\mlt}$=$1.63).   

Since the RS Canum Venaticorum stars show activity,  mass transfer episodes 
between active regions of both components are possible. 
In this case, the application of the single-star theoretical models 
used by Popper 
may not be relevant since the evolution of both components would not be 
independent any longer.  
However, for UV Psc, mass transfer does not account for the age difference because the 
computation of the Roche lobe radii (e.g. with the formulae given by
 Eggleton 1983) 
shows that none of these stars have overflown their Roche lobes.  
Therefore, we consider that UV Psc is a well-detached system and assume that 
the evolution of each component is independent. 

Another concern about RS CVn stars is the possible distortion of 
light curves due to starspots, which could bias the derivation of their
 fundamental stellar parameters. 
Jassur \& Kermani (1994) suggested the presence of cool spots on the 
surface of the primary component, but according to Popper (1997, and  
references therein), this should not affect the value of the quantities 
used in this paper. The period has been recently revised by Sowell et 
al. (2001) who didn't find any evidence for a cyclic modulation or 
period change. Hence activity cannot account for the age discrepancy
either.

To tackle this problem, we explore the influence of each 
physical parameter. A stellar evolutionary code allowing  
to compute tracks for different and independent physical parameters 
is needed for such a purpose, and so we choose the CESAM code (Morel 1997) 
which fulfills this condition and which has  already been  
successfully tested in a variety of  
astrophysical contexts (e.g. Zahn {\al} 1997, Suran {\al} 2001, 
Cordier {\al} 2002). 

In this paper, 
we present a detailed study of the system UV Psc with the CESAM models. 
Our goal is twofold: 
1) to check if there exists a solution giving consistent ages for both
 components  for a set of ($Z$, $Y$), and  
2) to constrain {\mlt} for both components and check if these 
values are consistent with the generally assumed solar 
value, {\mlt}$_{\odot}$.  

The paper is organised as follows: Sect. 2 deals with the description of the 
CESAM  models used and the grid computed for the purpose of this paper. 
Sect. 3 presents the results which fit both components of the 
eclipsing binary UV Psc, with particular emphasis on the constraints 
derived from the {\teff}-radius diagram on the {\mlt} parameters. 
We also discuss the robustness of the results against changes
in the {\teff} scale and in diffusion.
Finally, Sect. 4 provides a summary of our results. 

\section{Evolutionary stellar models} 
\label{s:cesam}

The stellar evolution calculations were computed with the CESAM code
(Version 3, Morel \cite{morel97}), running at the Coimbra Observatory.
Details on the physics of these models can be found in
Lebreton {\al} (\cite{lebreton99}).
Here we just provide a short summary of the main CESAM physical inputs: 
the CEFF equation of state is used, including Coulomb corrections
to the pressure (Eggleton {\al} \cite{eggleton73},
Christensen-Dalsgaard \cite{jcd91});
the nuclear reactions rates are from Caughlan and Fowler (\cite{caughlan88});
the solar mixture is from Grevesse and Noels (\cite{grevesse93});
the OPAL opacities (Iglesias and Rogers \cite{iglesias96}) are used 
and complemented at low temperatures by opacity data from
Alexander and Ferguson (\cite{alexander93}) following a
prescription by Houdek \& Rogl (\cite{houdek96});
the atmosphere is described with an Eddington $T(\tau)$-law;
the convection is treated according to the mixing-length
theory from B\"ohm-Vitense (\cite{bohm-vitense58}),  
with the formalism of Henyey {\al} (1965),   
giving the mixing-length scale ({\mlt}${\times}H_p$) in terms of
a free parameter times the local pressure scale height $H_p$.

\begin{figure*}[htb] 
\centerline{\psfig{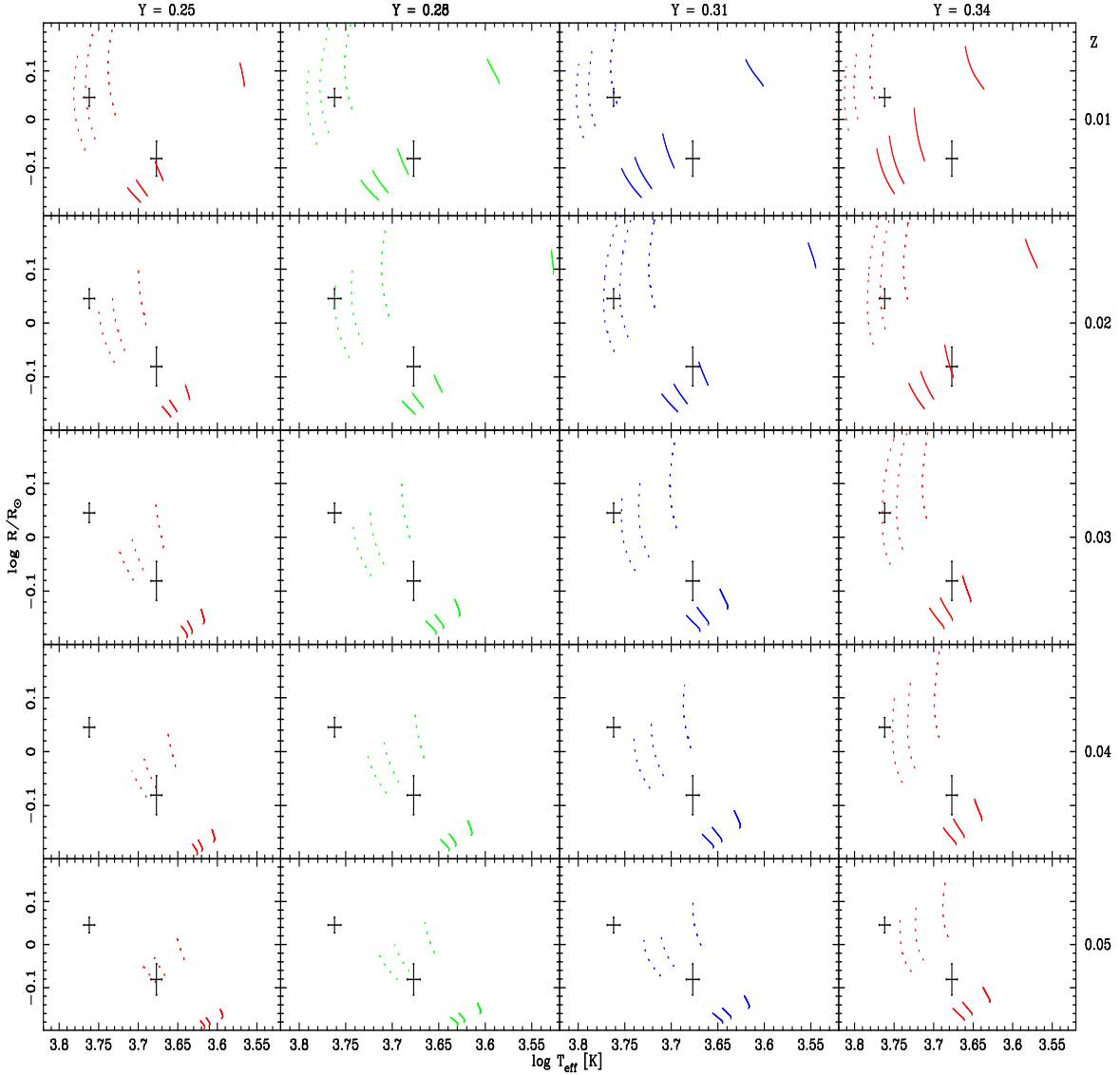}}
\caption{Location of UV Psc A and B (error bars in bold, the primary component being the hottest one) 
in the effective temperature-radius diagram with evolutionary tracks 
computed from the CESAM code 
for various chemical compositions. 
On each panel, tracks with the mass of the primary (dashed lines) and the 
secondary (solid lines) components are shown for 3 
different values of {\mlt} : {\mlt}$=$1.7, 1.2 and 0.8 
(from left to right).   
For $Z=$0.01 and $Z=$0.02, we also show {\mlt}$=$0.1 
(solid line at the coolest 
temperatures) for the secondary component.  
All tracks are computed from the ZAMS to the TAMS  or to 10 Gyr~(see text for details).
}
\label{fig:UVPsc}
\end{figure*}

With these physical ingredients, we computed a solar model that fits 
the observed luminosity and radius with an accuracy of $10^{-4}$,  
with {\mlt}${=}1.63 $,
helium abundance $Y_{\odot}$$=0.268$ and $Z_{\odot}$$=0.0175$
for the commonly accepted solar age of
$\sim$4.6 Gyr (Dziembowski {\al} \cite{dziembowski99})
and  solar abundances as derived by 
Grevesse \& Noels (\cite{grevesse93}).

Given this calibration, we computed a large grid of CESAM models from 
the Zero Age Main Sequence 
(ZAMS) to the terminal Age Main Sequence (TAMS, central hydrogen exhaustion), 
or for ages $t$ younger than 10 Gyr\footnote{UV Psc being a binary 
belonging to 
the close solar neighbourhood ($d\approx$~63 pc), ages older than the 
Galactic disc  
(likely to be 8-10 Gyr old, at most, see Carraro 2000) should be 
irrelevant here.}.    
This was done for each component of UV Psc  by fixing the masses 
to their accurately measured values: $M_A=$ 0.975 {\msun} 
and $M_B=$ 0.760 {\msun} (Popper 1997). 
This approximation is justified since the relative mass accuracy is 
better than 1\% in each star. \\
We considered different values of the metallicity $Z$ (0.01, 0.02, 0.03, 0.04  
and 0.05), helium content $Y$ (0.25, 0.28, 0.31 and 0.34) and mixing length 
parameter {\mlt} (0.8, 1.3 and 1.7) in a large three-dimensional
grid, since we do not want to introduce correlations between these
parameters. 
We also computed additional models for the secondary component 
with {\mlt}$=$0.1 for $Z=$ 0.01 and $Z=$ 0.02\footnote{
The internal structure of such low-{\mlt} models does not 
present any convective region at all (this can be explained by the fact
that the convective efficiency is proportional to the square of the mixing 
length and hence to $\alpha^{2}_{\rm MLT}$). The discussion on
 how realistic these 
models are is beyond the scope of this paper, but we would like to point out 
that we only used them to interpolate models with {\mlt} slightly lower 
than 0.8.}
We stress that the quantities $Z$ and $Y$ are independent,  
unlike most of the extant publicly available theoretical models 
where $Y$ is derived from a fixed law once $Z$ is determined. 
All the models computed with the input physics described above  
are displayed on Fig.~\ref{fig:UVPsc}.

\section{Analysis of the results} 
\label{s:results}

UV Psc is an eclipsing binary with very accurate masses and radii 
derived for both components. 
The fundamental data used in this paper (from Popper, 1997) 
are listed in Table~1 and are a key point of this study: the measures 
by Popper represent such a significant revision 
and improvement that a comparison with older data is irrelevant.  

For instance, Montesinos {\al} (1988) provided a comprehensive work 
on RS CVn EBs and derived a metal-rich solution for UV Psc ($Z=$0.04 and 
$Y=$0.25) from the log g-{\teff} diagram, but the data they used were 
very significantly revised by Popper ($\Delta$$M_A$$=$0.3 {\msun}, etc)
 hampering any useful comparison.  

\begin{table}[htb] 
\caption[]{Parameters from Popper (1997) for the primary (A) and secondary (B) 
components of UV Psc.}
\label{tab:mass}
\begin{flushleft}
\begin{center}  
\begin{tabular}{ccccc}
\hline 
\noalign{\smallskip}
Comp. & $M$           & $R$           & $\log$ ({\teff}/K) \\
      & [M$_{\odot}$] & [R$_{\odot}$] &                    \\
\noalign{\smallskip}
\hline 
\noalign{\smallskip}
A & 0.975$\pm$0.009  &  1.11$\pm$0.02  & 3.762$\pm$0.007    \\ 
B & 0.760$\pm$0.005   &  0.83$\pm$0.03  & 3.677$\pm$0.007    \\ 
\noalign{\smallskip}\hline
\end{tabular}
\end{center}
\end{flushleft}
\end{table}

First, before exploring all possible solutions in the  
($Z$, $Y$, {\mlt}, $t$) parameter space, interesting qualitative conclusions 
can be derived from a simple inspection of Fig.~1. 
Whatever the values of $Y$ and {\mlt} are, all  models with
 $Z$$\geq$0.04 
are  unable to match simultaneously the location of both components, 
strongly 
suggesting that metal-rich solutions can be ruled out.  
In addition, while a fit to both components seems to be possible in 
some panels --e.g., the upper left panel at $(Z,Y)=$(0.01,0.25)-- inconsistent 
ages appear, excluding these possible solutions. 
This is also the case for the solution found by Popper at $(Z,Y)=$(0.02,0.30)  
using the Geneva models. 

Altogether, this means that we expect to find solutions with consistent 
ages for chemical compositions in the range $Z=$0.01-0.02 and $Y=$0.28-0.34. 
At this level of qualitative discussion, it appears difficult to 
infer strong conclusions about the {\mlt} parameter but the panels of Fig.~1 
suggest that {\mltb} may be systematically smaller that {\mlta}.    

To obtain more detailed conclusions, a systematic work matching some selected 
relevant criteria has to be performed.   
For this purpose, we explore by 
interpolation\footnote{$Z$, $Y$ and {\mlt} in steps of 0.001, 0.005  
and 0.05, respectively.} the 
($Z$, $Y$, {\mlt}, $t$) parameter space described in 
the previous section to find out the best mathematical solutions 
matching 
the location of UV Psc in the {\teff}-radius diagram for both components 
($i=$ A, B). This is performed by minimizing the 
$\chi^2$-functional 
\beqa
\chi^2 (Z_i, Y_i, \alpha_{\rm MLT \it i}, t_i)  & = &  
\left(\frac{\log T_{\rm eff}(track) - \log T_{{\rm eff } i}}{\sigma(\log T_{{\rm eff }  i})}\right)^2 \nonumber \\
 & & +  
\left(\frac{\log R(track) - \log R_i}{\sigma(\log R_i)}\right)^2  
\eeqa
We assume that both components share the same chemical composition, i.e. 
$Z_A = Z_B$ and  $Y_A = Y_B$. 
Moreover, they should also have equal (or at least consistent) ages, hence
$\| t_A - t_B \|$ should be minimal. 

\begin{figure}[htb] 
\centerline{\psfig{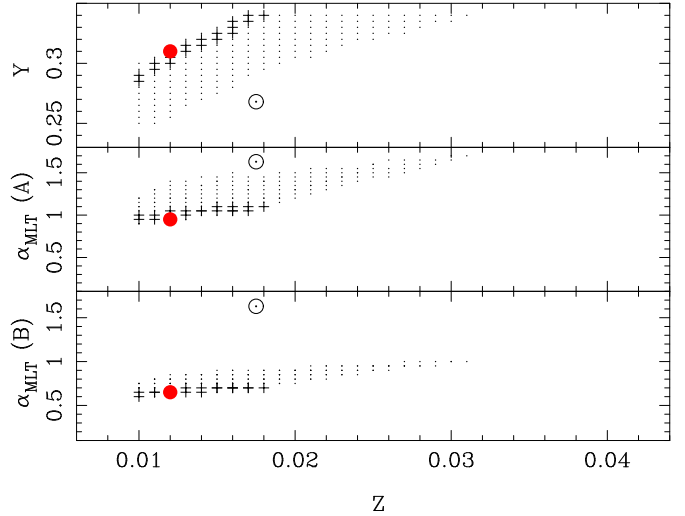}}
\caption{($Z$, $Y$, {\mlta}, {\mltb}) solutions (dots) matching the location 
of both UV Psc A and B in the {\teff}-radius diagram ({\teff}s from 
Popper, 1997) with the CESAM evolutionary tracks. 
Adding a constraint to obtain consistent ages for both components 
($\Delta$$t$/$t$$\leq$20\%) provides the subsample shown as crosses. 
The best agreement in age (0.9\%) is shown with a filled circle. 
The Sun (dotted circle) is also shown for comparison.     
}
\label{fig:popper}
\end{figure}

In Fig.~\ref{fig:popper}, we show all the solutions in three 
projections: $Z$-$Y$, $Z-${\mlta} and $Z-${\mltb} diagrams. 
The solutions (small dots on Fig.~\ref{fig:popper}) which have a 
good $\chi^2$ 
for both components 
($\chi$$^{2}_{i}$ $\leq$ $\chi$$^{2}_{i,min}$ $+$ 1 with $i=$A, B) 
show that 
i) high metallicities ($Z>$0.032) are indeed ruled out, 
ii) 1.7 $\geq${\mlta} $\geq$ 0.9, and 
iii) 1.0 $\geq${\mltb} $\geq$ 0.6. \\

\begin{figure}[htb]
\centerline{\psfig{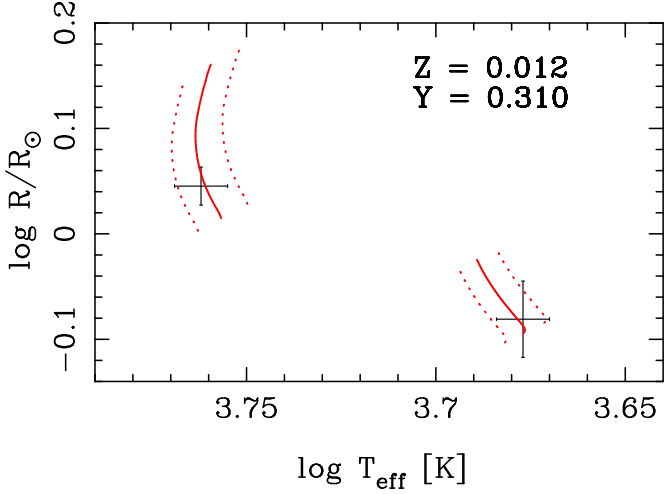}}
\caption{Best fit solution (filled circle in Fig. 2) for both UV Psc A and B 
in the temperature-radius diagram ({\teff}s and radii from Popper, 1997) 
with the evolutionary tracks computed 
from the CESAM code: {\mlta}$=$0.95 $\pm$0.12 and {\mltb}$=$0.65 $\pm$0.07. 
}
\label{fig:fit}
\end{figure}

Adding the constraint that the ages are consistent with each other  
($\Delta$t/t$\leq$20\%) provides the subsample shown with crosses. 
It defines a much smaller region (Fig.~\ref{fig:popper}) and shows 
that 
i) the metallicity is solar or sub-solar ($Z<$0.019) and the helium 
content is larger than the solar value ($Y>$0.28), 
ii) 1.1 $\geq$ {\mlta} $\geq$ 0.9, and 
iii) 0.7 $\geq$ {\mltb} $\geq$ 0.6. \\

The best solution (filled circle in Fig.~\ref{fig:popper}) is obtained with 
a $Z=$0.012, $Y=$0.31 chemical composition and provides an excellent 
agreement in age (0.9\% discrepancy).  
We predict an age of about 1.9 Gyr  
for both components of UV Psc A (1.92$^{+0.6}_{-0.4}$ Gyr) 
and B (1.94$^{+3.0}_{-0.4}$ Gyr). 
While a unique ($Z$, $Y$, $t$) combination is needed to simultaneously 
match the (masses, radii, {\teff}s) of both components, 
this solution implies a different value for the convection parameter for the 
two components of the binary:  {\mlt}(UV Psc A) $=$ 0.95$\pm$0.12 and 
{\mlt}(UV Psc B) $=$ 0.65$\pm$0.07. 
The fit provided by  this solution is shown in Fig.~\ref{fig:fit}.

The non-solar values for {\mlt} parameters that we derive obviously deserve
 further comments, 
but the robustness of these solutions needs to be discussed first.

\subsection{Influence of the {\teff} scaling}
\label{section:mlt}

Since our main concern was to solve the age discrepancy found by 
Popper (1997), 
we used the same values for the mass, radius and {\teff}.
We have shown in the previous section that this discrepancy can be solved 
within the same data set. However, 
whilst the individual masses and radii are very accurately 
defined, the {\teff}s are not direct quantities, and depend on  colour 
indices and photometric calibrations.
We would like to point out that according Popper (\cite{popper1998})
the presence of dark spots has relatively little effect on the color index
of a star.
To test the possible influence of the {\teff} scaling on our results, 
we used as an alternative the metallicity-dependent empirical calibrations 
of Alonso {\al} (\cite{alonso96}), {\teff} $=$f($B-V$, [Fe/H]) 
and used the $B-V$ indices of both components from Table 6 of Popper (1997)  
assuming that the reddening can be neglected because UV Psc is in the inner  
solar neighbourhood ($\sim$ 63 pc according to its Hipparcos parallax).  
%
%
Since there is no direct determination of the iron abundance [Fe/H] 
available for 
UV Psc, we have computed both {\teff}s from the Alonso {\al}'s formula (1) 
assuming three different metallicities: 
{\teffa}$=$ 5574 K and {\teffb}$=$ 4506 K ($Z=$ 0.01),
{\teffa}$=$ 5682 K and {\teffb}$=$ 4586 K ($Z=$ 0.02),
{\teffa}$=$ 5754 K and {\teffb}$=$ 4638 K ($Z=$ 0.03). 
The uncertainty on the primary (secondary) component is less than
 150 K (100 K), and 
so -- to be conservative -- we consider these upper limits as the uncertainty. 
For further comparisons, we take the {\teff}s derived for $Z=$0.02. 
They are cooler by about 100 K (170 K) for the primary (secondary) component 
in comparison to the {\teff}s that we have used previously (see Table 1).  

%
%

\begin{figure}[htb]
\centerline{\psfig{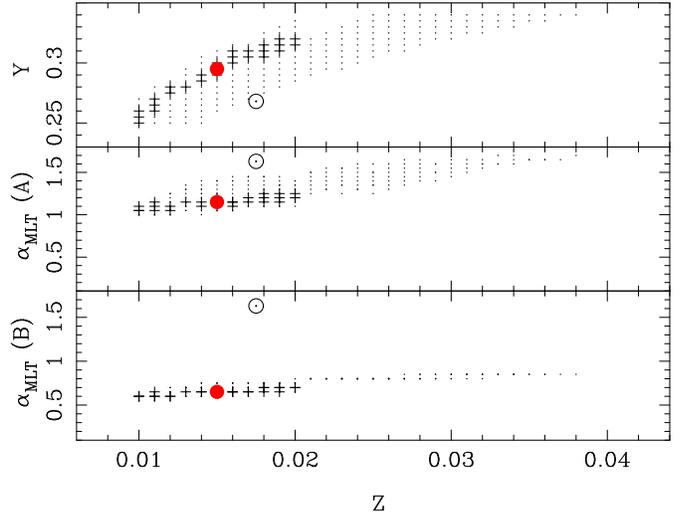}}
\caption{Same as Fig.~\ref{fig:popper} but with the {\teff}s derived from 
the Alonso {\al} (1996) calibrations. 
The best agreement in age (1.6\%) is shown as a filled circle.
}
\label{fig:alonso}
\end{figure}

The influence of adopting the {\teff}s we have derived from the Alonso {\al} 
calibrations can be seen in Fig.~\ref{fig:alonso}. 
A comparison with Fig.~\ref{fig:popper} shows that the global solutions 
(dots) are fully consistent with each other. 
However, the best solution according to the criteria defined in Section 3 
is slightly different: ($Z$, $Y$) $=$ (0.015, 0.295), the ages predicted being 
about twice older (UV Psc A: $\sim$ 4.07 Gyr and B: $\sim$ 4.14 Gyr). 
In spite of these shifts, we note that a different value 
is still needed for the convection parameter of the two components 
({\mlta}$=$ 1.15 and {\mltb}$=$ 0.65).    
The difference $\Delta${\mlt} is even increased 
in comparison to the results inferred from the Popper (1997)'s {\teff} scale,
and  
hence the conclusions of the previous sub-section remain qualitatively 
unchanged.  Quantitatively, a systematic error of about +0.20 in \mlta ~(and
a negligible one in \mltb) 
may be due to the uncertainties in the temperature scaling. 

\subsection{Influence of diffusion}
\label{sec:diffusion}

\begin{figure}[htb]
\centerline{\psfig{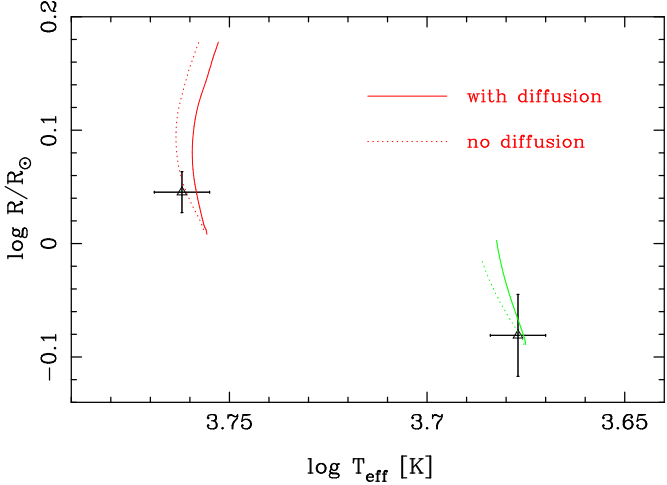}}
\caption{The effect of including microscopic diffusion models, 
assuming the same composition (Z,Y)=(0.012,0.31) for each component.
}
\label{fig:diffusion}
\end{figure}

Another potential source of systematic error is the influence of
diffusion. We have treated this by computing CESAM models incorporating
the microscopic diffusion as described by the simplified formalism
proposed by Michaud \& Proffitt (1993) with metal elements as
trace elements. Note that the radiative
acceleration can be neglected for sub-solar stars (Turcotte et al., 1998).
As shown on Fig.~\ref{fig:diffusion}, diffusion almost mimics an
increase in \mlt, even though the slopes of the tracks are
obviously changed. At fixed radius, the effect of diffusion is
to change the effective temperatures by some 65 K (55 K), well within
the 1-$\sigma$ uncertainties in \teff ~discussed above. Alternatively,
taking these uncertainties at face value, one would have to increase
\mlt by about 0.11 (0.07) at most.

In summary, even though the absolute values of \mlt ~may change due to
these combined effects, the trend of increased \mlt with mass remains robust.

\subsection{Universality of the {\mlt} parameter ?}
\label{sec:mlt}

We have quantitatively confirmed the suggestion made by  Clausen {\al} (1999)
that the adoption of a significantly smaller 
{\mlt} for the secondary component may remove (or at least 
decrease) the general problem of age discrepancy presented in the Introduction. 
While it is obvious that only observational constraints on the 
 chemical composition of UV Psc would allow a proper test, 
it seems that this problem can be solved theoretically for this binary. 
The solution obtained for this system 
suggests that 
i) the components have a different {\mlt} ($\Delta${\mlt}$=$0.3$\pm$0.14) 
and 
ii) these values are clearly different from the solar one. 

This implies that the solar value of the {\mlt} is not universal, 
an hypothesis made in most of 
the currently available stellar evolutionary models.  

\begin{figure}[htb]
\centerline{\psfig{file=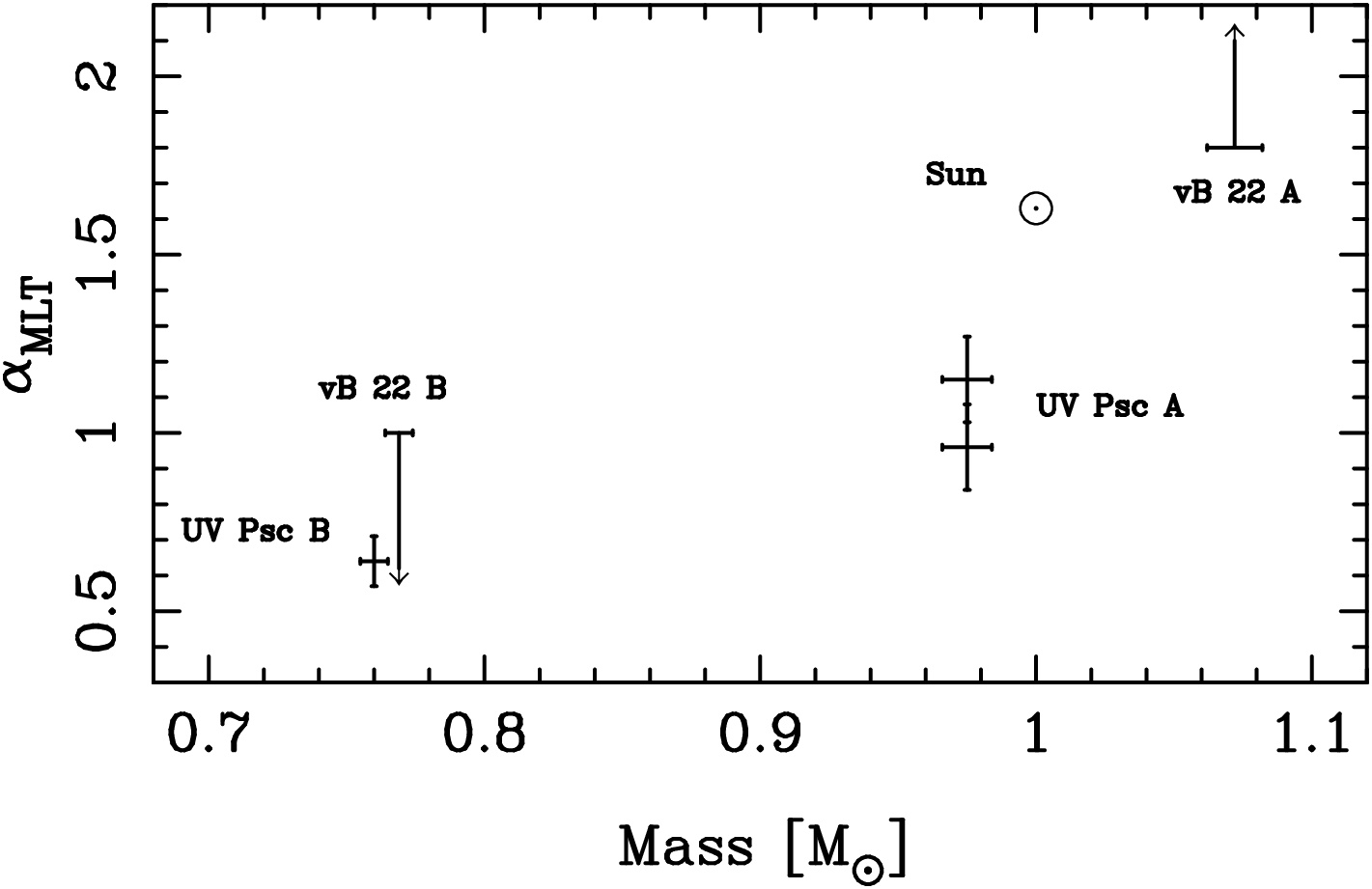,width=\columnwidth,angle=-90.}}
\caption{Best fit solution for both UV Psc A and B 
in a mass-{\mlt} diagram. The value of {\mlta} is sligthly different 
according to the assumed {\teff}s: {\mlt}$=$0.95 (Popper 1997), 
{\mlt}$=$1.15 (Alonso {\al} 1996, see text). 
The solution is independent of the choice of 
the {\teff}s for the secondary component. 
The Sun (cf. \S2) is shown for comparison, as well as estimates 
(shown as upper and lower limits) derived from CESAM models 
with the same input physics for both components of 
the Hyades binary vB 22 (Lebreton {\al} 2001). 
}
\label{fig:mlt_mass}
\end{figure}

As for the trend with mass, a recent study on the Hyades 
eclipsing binary vB22 (Lebreton {\al} 2001) 
seems to support it. We show in Fig.~\ref{fig:mlt_mass} that UV Psc 
and vB22 are clearly not consistent with the solar {\mlt} value. 
Moreover, Lebreton {\al} (2001) show that the slope of the main-sequence in 
the Hyades suggests that {\mlt} could decrease with mass, in particular they 
obtained {\mlt}$\leq$1.4 below 1 {\msun}. The results we have obtained 
for UV Psc give additional support to this trend. 
However we note that
this trend seems to contradict the results of Ludwig et al. (1999) coming
from detailed hydrodynamical simulations where, for a fixed gravity, {\mlt} 
increases with deacreasing \teff.


More results are needed before reaching any definitive conclusion, because 
a larger sample may well reveal a scatter of {\mlt} for a given mass. 
For instance, recent calibrations of {\mlt} in visual binary stars were 
performed by P. Morel and collaborators using the CESAM code 
(with different physical ingredients than those  used in this paper).  
For the stars in the range of mass relevant in this paper, they find   
$\alpha$ Cen B, $M=$ 0.97 {\msun}, {\mlt}$=$ 1.53 (Morel et al 2000a),  
$\iota$ Peg B, $M=$0.81 {\msun}, {\mlt}$=$1.36 (Morel et al 2000b) 
and 85 Peg A, $M=$0.84 {\msun}, {\mlt}$=$1.80 (Fernandes {\al} 2002). 
Since both $\iota$ Peg B and 85 Peg A have very similar masses, this seems to 
support the idea of a dispersion of {\mlt} at a given mass. 
Solar models with and without diffusion yield different values in the
internal convective regions (1.7 vs 1.9, Richard {\al} 1996). 
There are no further data for main sequence stars with larger masses. Studies of  more
massive stars have dealt exclusively with red giant stars. For instance,
{\mlt} seems to increase to values up to 3
to reproduce the red giant branches in open clusters (Stothers \& Chin,
1997), while 
Iwamoto \& Saio (1999) present a trend of {\mlt} increasing with 
mass and  metallicity again for giant stars. In these stars 
overshooting is playing a major role and so disentangling the
relative effects of mass, metallicity and overshooting is far
from trivial.

 Therefore, at this stage, a relation between mass and {\mlt} along
the main sequence remains highly 
premature and must be seen with caution. 

We also note, in passing, that our results are not in contradiction with
the recent analysis by Palmieri {\al} (2002), who on the basis of the
colours of red giants in globular clusters concluded that there is no
dependence of {\mlt} with metallicity. The extremely narrow range in mass
probed by these red giants would be nearly equivalent to a single point on
our diagram. Since the detailed modelling of red giants is more complicated
than the one of main sequence stars, a larger sample of double-lined,
eclipsing binaries with components in the 0.7-1.1 \msun would offer a much
cleaner and definitive test.

In any case, all these possible variations and their trends reflect 
a shortcoming of the MLT theory more than a real physical effect. If
one insists in using MLT, its parameter {\mlt} should be varied. This
is perhaps the first time that such a result is obtained using observations,
since the variablity of the effective {\mlt} is a well-know result of detailed
hydrodynamical simulations of stellar convection (Ludwig et al. 1999). 

On the other hand we note that the value of {\mlt} determined 
in atmosphere of the stars gives in general
different results in relation to those obtained from the stellar interior analysis. For instance, 
van't Veer-Menneret \& M\'egessier (1996) found that the spectral fitting of the first
for Balmer lines of the solar spectrum required {\mlt}$=0.5$, a result
confirmed by Fuhrmann {\al} (1993) for other cool stars, while Gardiner,
Kupka \& Smalley (1999) found that {\mlt} $\sim 0.5$ for {\teff}s below 6000 K
but {\mlt} $\geq 1.25$ in the range 6000--7000 K to possibly decrease
again to {\mlt} $ = 0.5$ above 7000 K.

\subsection{Implications for the helium to metal ratio}
\label{sec:dydz}

Since we have derived $Z$ and $Y$ independently for UV Psc, 
the helium to metal ratio ($\Delta$$Y$/$\Delta$$Z$)  
of this system can be inferred assuming a primordial helium abundance. 
The initial helium content is fixed by taking the currently assumed  
law $Y=$ $Y_p$ $+$ ($\Delta$$Y$/$\Delta$$Z$) $\times$ $Z$, where $Y_p$ is 
the primordial helium abundance. 
If we consider $Y_p=$0.235 (Peimbert {\al} 2000) and use the ($Z$, $Y$) solution 
that we have found for the UV Psc binary, we derive 
$\Delta$$Y$/$\Delta$$Z$ $\sim$ 6 and 4 by using the Popper (1997) or Alonso {\al} 
(1996) {\teff}s, respectively.  
In both cases, the ratio is large in comparison to what can be obtained from 
the solar model described previously: 
($\Delta$$Y$/$\Delta$$Z$)$_{\odot}$ $\sim$ 1.9. 
This may suggest that the currently assumed simple linear relationship 
between $Z$ and $Y$ is not universal and/or that $\Delta$$Y$/$\Delta$$Z$ 
is not unique. \\
The possible relation between $Y$ and $Z$ is one of the most interesting 
and puzzling problems in astrophysics since the pioneering work by 
Faulkner (1967). 

The most recent observations in HII regions and planetary nebulae as well as 
semi-empical determinations using theoretical stellar models - including the 
detailed calibration of the Sun - (cf. Ribas {\al} 2000, Fernandes 2001 and references therein) 
support values around 2 or 3 
, assuming $Y_p$ $\sim$ 0.23. 
Nevertheless, some exceptions indicate values larger than 4   
(e.g. Pagel {\al} 1992 from observations of extragalactic HII regions, 
and Belikov {\al} 1998 from the main sequence of the Pleiades). 
Our determination clearly supports the idea of a large $\Delta$$Y$/$\Delta$$Z$ 
ratio. \\

\section{Conclusions} 

We study the eclipsing binary UV Psc, which is of particular 
interest since Popper (1997) derived age estimates for both components 
which differed by some 11 Gyr.  
Adopting his same data set (masses, radii and {\teff}s)
 we compute a large grid of CESAM models for each 
component\footnote{These models are available upon request 
(email: {\tt jmfernan@mat.uc.pt}).}.  
We obtain a best fit in the {\teff}-radius diagram  
for a common chemical composition ($Z$, $Y$)$=$(0.012, 0.31), 
a negligible age difference, 
and a different MLT  
parameter ({\mlta}$=$ 0.95$\pm$0.12 and {\mltb}$=$ 0.65$\pm$0.07). 
The difference {\mlta} $-$ {\mltb} $=$ 0.3 is significant and remains
robust against changes in temperature calibrations and the
inclusion of diffusion.

Within the framework of the MLT theory, 
this suggests that fixing {\mlt} to its solar value ($\sim$1.6), the usual 
hypothesis made by most  stellar evolutionary models, may not be correct. 
Secondly, since the {\mlt} is smaller for the less massive component, this 
suggests that the {\mlt} parameter may decrease with the stellar mass. 
This trend needs further confirmation with other binary stars 
with accurate data. Again, this may reflect a problem of the application
of the MLT theory, more than a real physical effect, but it has to be
taken into account when using MLT. On the other hand as the main observational errors come from the \teff, 
new and independent determinations are needed for both stars.

So far, few other binaries have data with the precision required to
confirm this trend and observational programs are highly needed 
(e.g. Kurpinska-Winiarska \& Oblak, 2000). 

\begin{acknowledgements}
We warmly thank P. Morel for making available CESAM, a superb tool for
stellar evolution, and without which this study could not have been
performed.
EL is supported by a "Funda{\c{c}}\~ao para a  Ci\^encia e Tecnologia'' 
(FCT) postdoctoral fellowship (grant SFRH/BPD/5556/2001).  
This work was partially 
supported by the project "PESO/P/PRO/15128/1999" from the FCT, and
by {"Conv\'enio Embaixada de Fran{\c{c}}a-ICCTI (59-B0)"}.  
This research has made use of the SIMBAD database operated at CDS, Strasbourg, France, 
and of NASA's Astrophysics Data System Abstract Service. 
\end{acknowledgements}


\end{document}